\documentclass[twocolumn]{aastex61}
\pdfoutput=1 %for arXiv submission
\usepackage{amsmath,amstext}
\usepackage[T1]{fontenc}
\usepackage{apjfonts} 
\usepackage[figure,figure*]{hypcap}

\newcommand{\kms}{km s$^{-1}$}
\newcommand{\hst}{{\it HST}}
\newcommand{\ndash}{-}
\shorttitle{Distance to NGC 4993}
\shortauthors{Hjorth et al.}

\begin{document}

\title{The Distance to NGC 4993: The Host Galaxy of the Gravitational-wave Event 
GW170817}
\author{Jens Hjorth}
\affiliation{Dark Cosmology Centre, Niels Bohr Institute, University of Copenhagen, 
Juliane Maries Vej 30, DK-2100 Copenhagen \O, Denmark}
\author{Andrew J. Levan}
\affiliation{Department of Physics, University of Warwick, Coventry, CV4 7AL, UK}
\author{Nial R. Tanvir}
\affiliation{Department of Physics and Astronomy, University of Leicester, LE1 7RH, UK}
\author{Joe D. Lyman}
\affiliation{Department of Physics, University of Warwick, Coventry, CV4 7AL, UK}
\author{Rados{\l}aw Wojtak}
\affiliation{Dark Cosmology Centre, Niels Bohr Institute, University of Copenhagen, 
Juliane Maries Vej 30, DK-2100 Copenhagen \O, Denmark}
\author{Sophie L. Schr\o der}
\affiliation{Dark Cosmology Centre, Niels Bohr Institute, University of Copenhagen, 
Juliane Maries Vej 30, DK-2100 Copenhagen \O, Denmark}
\author{Ilya Mandel}
\affiliation{Birmingham Institute for Gravitational Wave Astronomy and School of 
Physics and Astronomy, University of Birmingham, Birmingham, B15 2TT, UK}
\author{Christa Gall}
\affiliation{Dark Cosmology Centre, Niels Bohr Institute, University of Copenhagen, 
Juliane Maries Vej 30, DK-2100 Copenhagen \O, Denmark}
\author{Sofie H. Bruun}
\affiliation{Dark Cosmology Centre, Niels Bohr Institute, University of Copenhagen, 
Juliane Maries Vej 30, DK-2100 Copenhagen \O, Denmark}

\received{2017 September 29}
\revised{revised 2017 October 2}
\accepted{2017 October 3; published 2017 October 16}

\begin{abstract}
The historic detection of gravitational waves from a binary neutron star merger 
(GW170817) and its electromagnetic counterpart led to the first accurate 
(sub-arcsecond) localization of a gravitational-wave event. The transient was found 
to be $\sim$10\arcsec\ from the nucleus of the S0 galaxy NGC\,4993. We report here 
the luminosity distance to this galaxy using two independent methods.
(1) Based on our MUSE/VLT measurement of the heliocentric redshift 
($z_{\rm helio}=0.009783\pm0.000023$) we infer the systemic recession velocity of 
the NGC\,4993 group of galaxies in the cosmic microwave background (CMB) frame to be 
$v_{\rm CMB}=3231 \pm 53$ \kms. Using constrained cosmological 
simulations we estimate the line-of-sight peculiar velocity to be 
$v_{\rm pec}=307 \pm 230$ \kms, resulting in a cosmic velocity of
$v_{\rm cosmic}=2924 \pm 236$ \kms\ ($z_{\rm cosmic}=0.00980\pm 0.00079$) 
and a distance of $D_z=40.4\pm 3.4$ Mpc assuming a local Hubble constant of 
$H_0=73.24\pm 1.74$ \kms Mpc$^{-1}$. 
(2) Using {\it Hubble Space Telescope} measurements of the effective radius 
($15\farcs 5\pm 1\farcs5$)
and contained intensity and MUSE/VLT measurements of the velocity dispersion, 
we place NGC 4993 on the Fundamental Plane (FP) of E and S0 galaxies. Comparing to a 
frame of 10 clusters containing 226 galaxies, this yields a distance estimate of 
$D_{\rm FP}=44.0\pm 7.5$ Mpc. The combined redshift and FP distance is 
$D_{\rm NGC 4993}= 41.0\pm 3.1$ Mpc. This `electromagnetic' distance estimate 
is consistent with the independent measurement of the distance to GW170817 as 
obtained from the gravitational-wave signal ($D_{\rm GW}= 43.8^{+2.9}_{-6.9}$ Mpc) 
and confirms that GW170817 occurred in NGC 4993.
\end{abstract}


\section{Introduction}

GW170817 was the first gravitational-wave event arising from a binary neutron 
star (NS) merger to have been detected by LIGO/Virgo \citep{lvcdisc17}. 
The source was localized to a sky region of 28 deg$^2$ purely using 
gravitational-wave data from the three interferometers. As such, 
it provided the first realistic chance of detecting an electromagnetic counterpart, 
as outlined in \citet{lvccapstone}: 2 s after the merger, {\it Fermi} and 
{\it INTEGRAL} detected a weak gamma-ray burst, and half a day after the event, an 
optical \citep{coulter17} and near-infrared \citep[(NIR)][]{tanvir17} counterpart was
localized to sub-arcsecond precision, $\sim 10\arcsec$ from the nucleus of the 
S0 galaxy NGC 4993 \citep{lvccapstone}. It exhibited an unprecedented optical/NIR 
lightcurve and spectral evolution \citep{lvccapstone,tanvir17}, strongly suggestive 
of formation of very heavy elements (lanthanides) out of the tidally ejected 
neutron-rich material in the merger \citep{pian17,tanvir17}.

It is highly likely that the neutron star--neutron star (NS-NS) merger occurred in NGC 4993. 
A precise distance to NGC\,4993 is therefore required in order to understand the energetics 
of the event. Furthermore, since NS--NS mergers provide a route to $H_0$ via the direct 
``standard siren'' measurement provided by the LIGO and Virgo Collaborations
\citep{lvchubble}, it is valuable to have an accurate redshift as well as independent 
electromagnetic distance estimates. The redshift, especially if corrected to the Hubble 
flow, is vital in constructing the gravitational-wave Hubble diagram, while an independent 
distance estimate can confirm the galaxy association.

To our knowledge, there are no direct measurements of the distance to NGC\,4993 
\citep{2009AJ....138..323T}. NGC 4993 belongs to a group of galaxies (see 
Section 3.2). Two galaxies have distances in the Cosmicflows-3 data release 
\citep{2016AJ....152...50T}. NGC 4970 (PGC 45466) has a Fundamental 
Plane (FP) distance of $39.8\pm 10.3$ Mpc from the 6dF Galaxy Survey 
\citep{2014MNRAS.445.2677S} while ESO508-G019 (PGC 45666) has a 
Tully--Fisher distance of $37.5\pm 5.6$ Mpc from $I$-band and {\it Spitzer} 
photometry \citep{1997ApJS..109..333W,2009ApJS..182..474S,2014MNRAS.444..527S}.
Combined, these estimates suggest a group distance of $38.0\pm 4.9$ Mpc.

Section 2 summarizes the origin of the data used to provide new distance estimates 
to NGC 4993. In Section 3 we provide an updated distance along this avenue. We use 
a new redshift and estimate of the peculiar velocity of NGC\,4993. In Section 4, 
we obtain the first direct estimate of the distance to NGC 4993 itself using the 
FP method. We discuss the results in Section 5 and summarize in Section 6.

\section{Data}

The data used in this Letter are described in \citet{levan17}. NGC\,4993 was observed 
with the Very Large Telescope (VLT) and the MUSE Integral Field Spectrograph on 2017 August 18.

We also used an Advanced Camera for Surveys F606W {Hubble Space Telescope} (\hst) 
image of NGC\,4993 obtained on 2017 April 28. The image was reduced via 
{\tt astrodrizzle}, with the final scale set to 0\farcs07.

\section{Redshift Distance}
\subsection{Redshift of NGC 4993}

An updated value of the heliocentric redshift of NGC\,4993 is obtained from the
MUSE/VLT observation reported by \citet{levan17}. Based on fits to the absorption 
lines from stars in the center of the galaxy, the heliocentric recession velocity 
is determined to be $v_{\rm helio}=2933\pm 7$ \kms\ corresponding to a heliocentric 
redshift of $z_{\rm helio} = 0.009783\pm 0.000023$. 

\subsection{Redshift of the NGC 4993 Group of Galaxies}
\begin{figure*}[h]
    \centering
    \includegraphics[width=1.0\hsize,angle=0]{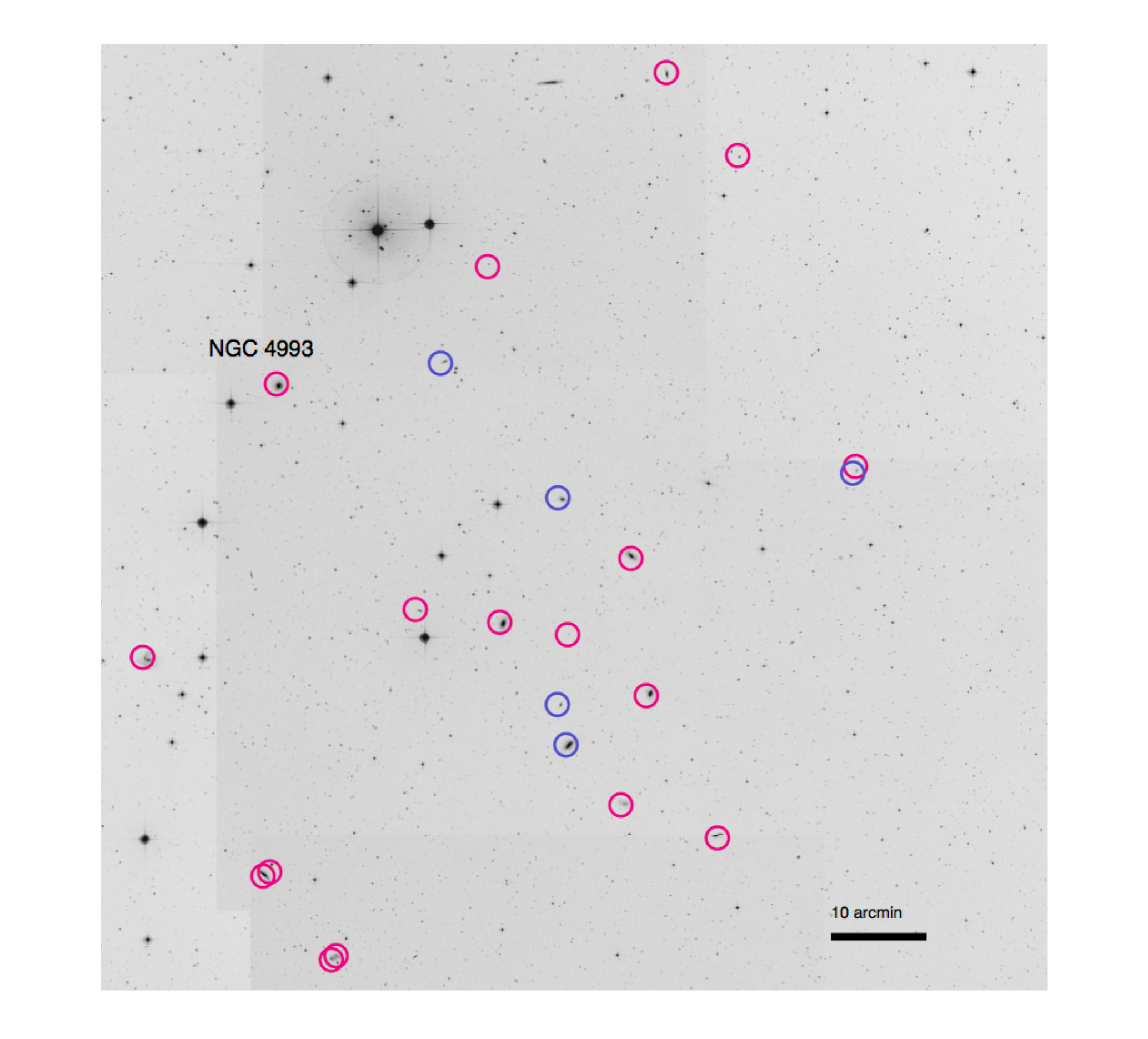}
\caption{Members of the NGC 4993 group of galaxies. Purple circles indicate
\citet{2017ApJ...843...16K} group galaxies with heliocentric velocities
less than 3005 \kms\ while blue circles indicate galaxies with
heliocentric velocities larger than 3169 \kms\ (see Figure 2).
This Digitized Sky Survey image is $2\,\deg$ on the side. 
North is up and East is to the left.}
\label{group}
\end{figure*}

\begin{figure}[h]
    \centering
    \includegraphics[width=8.8cm,angle=0]{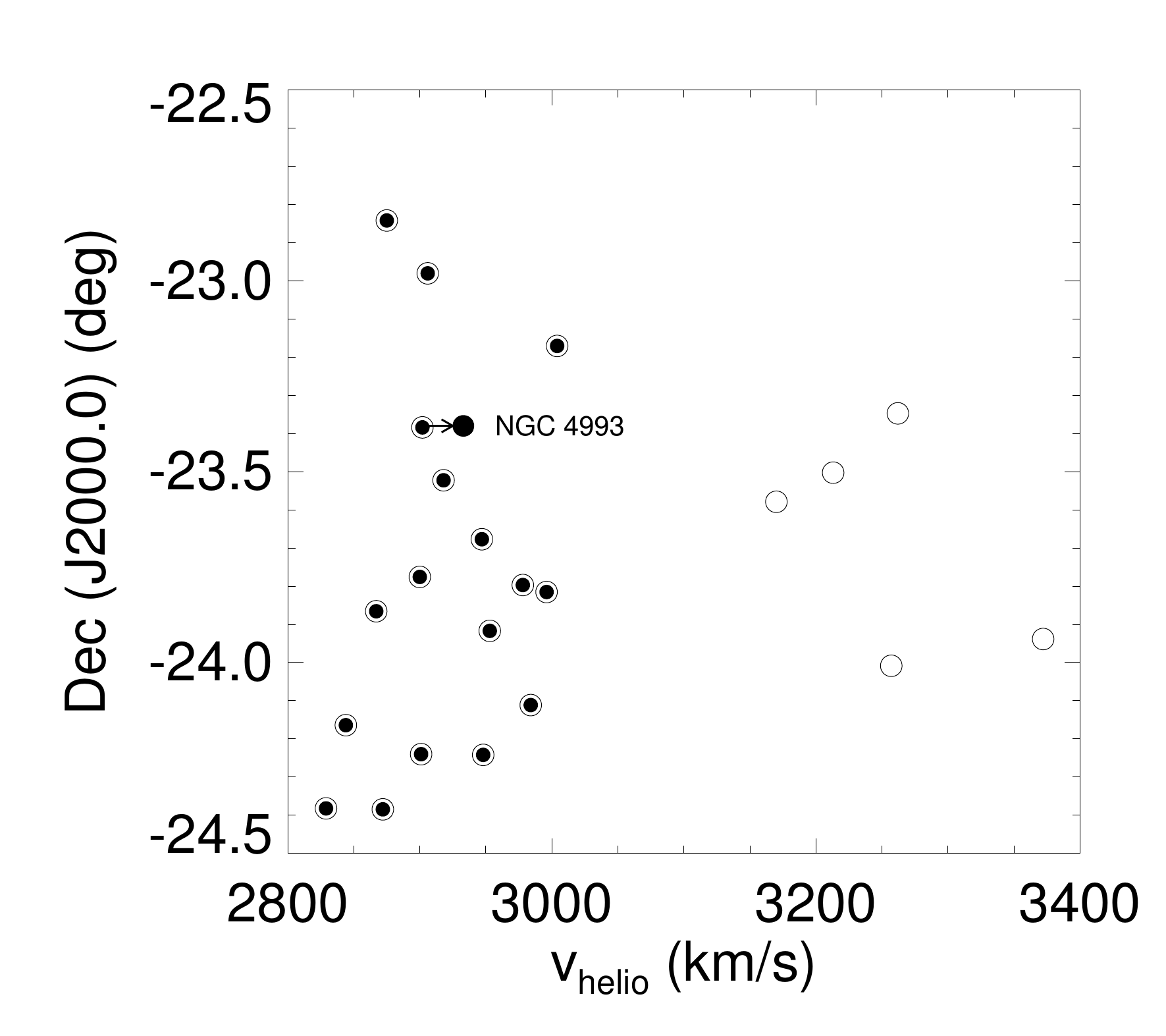}
\caption{Heliocentric velocities of the \citet{2017ApJ...843...16K} group 
PCG1 45466 containing NGC 4993. There is a clear gap in the heliocentric 
velocities, with no galaxies in the range 3005--3169 \kms, i.e., in the 
positive range above the mean velocity $v_{\rm helio} =$ 2995 \kms.
Considering only galaxies plotted as filled circles the mean velocity is
$v_{\rm helio} =$ 2921 \kms. The updated velocity of NGC 4993 is indicated.}
\label{group2}
\end{figure}

NGC\,4993 appears to be a member of a group of galaxies; see Figure~\ref{group}. 
\citet{2007ApJ...655..790C} list it to be one of 46 members of a group of galaxies 
in the 2MASS redshift survey (group number 955, with an average heliocentric 
velocity of 2558 \kms\ and a velocity dispersion of 486.5 \kms). However, NGC 4993 
is located $\approx$4.2 Mpc in projection from the center of this putative group 
and a non-detection in X-rays appears to rule out such a rich relaxed group. 
\citet{2011MNRAS.412.2498M} find NGC 4993 to be one of 15 members with a velocity 
dispersion of 74 \kms. Recently, \citet{2017ApJ...843...16K} found it to be 
one of 22 members of a group with a mean heliocentric velocity of 2995 \kms\ 
and a velocity dispersion of 118 \kms\ (our calculation of the rms velocity spread
of the 22 galaxies is 155 \kms). However, it is evident from Figure~\ref{group2}
that five of the galaxies have a much higher recession velocity than the rest
and so do not appear to be part of a relaxed group. We therefore exclude 
these 5 galaxies and consider the properties of the remaining 17 galaxies. 
We also update the NGC 4993 velocity from 2902 \kms\ to 2933 \kms\ (as measured 
here) and compute a mean heliocentric velocity of 2921 \kms\ and a velocity 
dispersion of 53 \kms. This small velocity dispersion and the
large extent of the structure suggest a relaxation time larger than $10^{10}$
years, and so the structure is unlikely to be a relaxed group. We therefore
adopt the velocity dispersion of the structure as the uncertainty in the recession
velocity (rather than the error in the mean, $53/\sqrt{16}=13$ \kms), i.e.,
$v_{\rm NGC 4993,group}= 2921 \pm 53$ \kms. This structure is shown in 
Figure~\ref{group}. Note that NGC 4993 is at the outskirts of the structure while 
its velocity is very close to the mean velocity of the structure.

Using the {\it WMAP} measurement of the cosmic microwave background (CMB)
dipole \citep{2009ApJS..180..225H}, we 
obtain\footnote{using the NED Velocity Correction Calculator at
https://ned.ipac.caltech.edu/forms/vel\_correction.html} a recession velocity 
in the frame defined by the CMB of $v_{\rm CMB}=3231 \pm 53$ \kms.

\subsection{Peculiar Velocity}
 \label{sub:pec}
 
To estimate the ``cosmic velocity", i.e., the recession velocity corresponding to 
pure Hubble flow, we need to take into account the  peculiar velocity due to 
large-scale structure:
\begin{equation}
v_{\rm cosmic}= v_{\rm CMB} - v_{\rm pec}.
\end{equation}
Following \citet{2014ApJ...796L...4L}, we make use of a dark-matter simulation 
from the Constrained Local Universe Simulations (CLUES) project to find the best 
estimate of the peculiar velocity of NGC 4993. The initial conditions of the 
simulation were generated using constraints from observations probing the 
distribution of galaxies and their peculiar velocities so that the final 
snapshot is a 3D representation of the observed local universe 
\citep[for technical details see][]{2010arXiv1005.2687G}. The simulation 
reproduces all main large-scale structures around the Local Group such as the 
Virgo cluster, the Coma cluster, the Great Attractor, and the Perseus--Pisces 
supercluster and the resulting peculiar velocity field. Structures on scales 
smaller than $\sim5 h^{-1} {\rm Mpc}$ are not subject to observational 
constraints and emerge from random sampling of the initial density field. 
Their evolution, however, is strongly influenced by tidal forces from the 
nearby large-scale structures. Therefore, the simulation provides a realistic 
and dynamically self-consistent model for the matter distribution and the 
velocity field in the local universe. The simulation adopted cosmological 
parameters from the third data release of the WMAP satellite (WMAP3), i.e., 
matter density $\Omega_{m}=0.24$, dimensionless Hubble parameter $h=0.73$ and 
normalization of the power spectrum $\sigma_{8}=0.76$. The simulation box has a 
side length equal to $160h^{-1} {\rm Mpc}$ and contains $1024^{3}$ particles.

\begin{figure}
\centerline{\includegraphics[width = 8.8 cm]{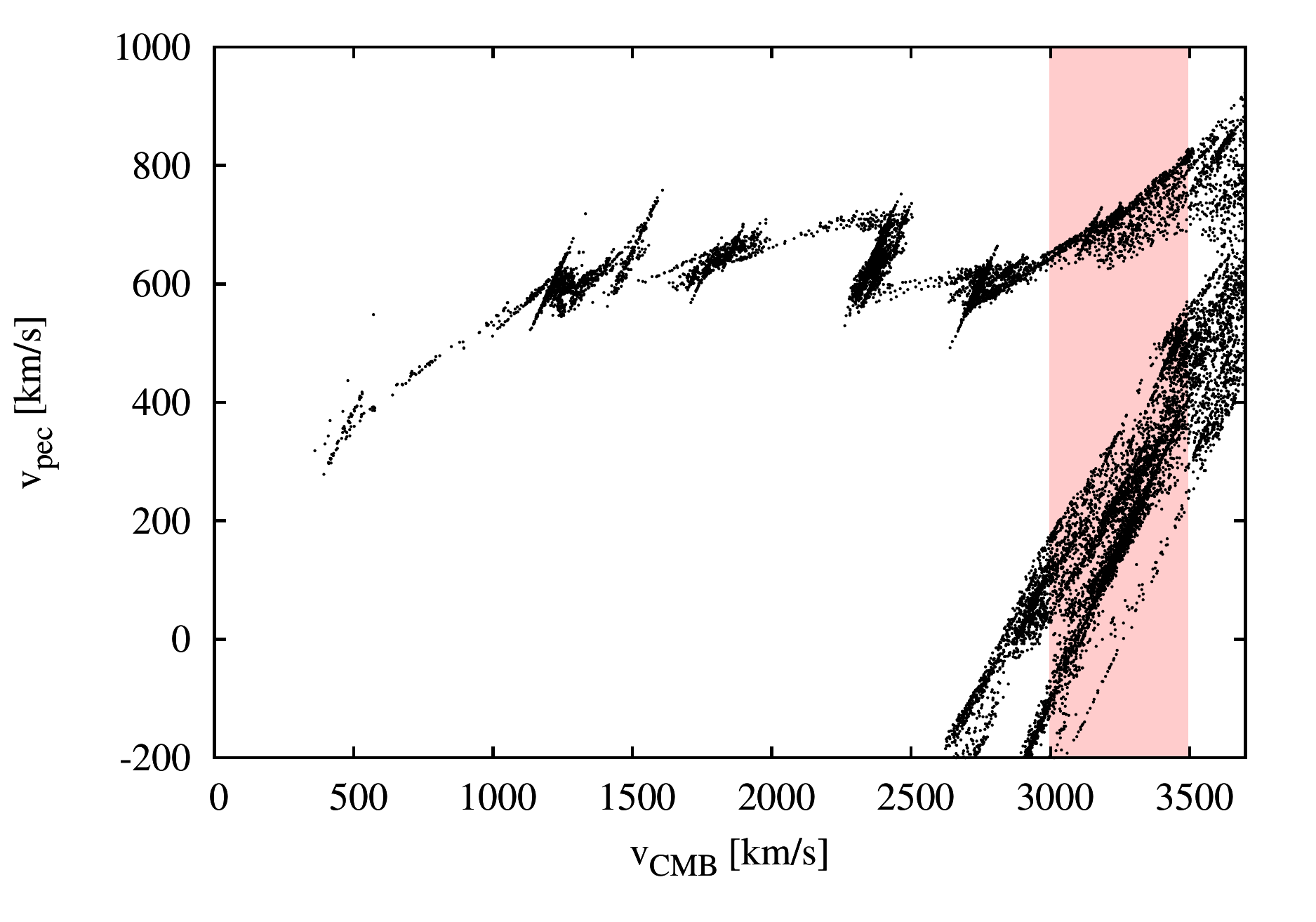}}
\caption{Peculiar velocities (line-of-sight component) of dark-matter 
particles in the constrained simulation, in the direction of NGC 4993. 
Velocity $v_{\rm CMB}$ is measured in the reference frame of the CMB 
with respect to an observer located at the Local Group (numerical analog). 
The red stripe indicates the location of NGC 4993.}
\label{pec}
\end{figure}

In order to find the position vector of the observational cone in the simulation 
box, we calculate the angular distances of NGC 4993 from three reference 
points that have unambiguous analogs in the numerical model: the direction of the 
peculiar velocity of the Local Group, the Great Attractor (Norma cluster),
and the Perseus cluster. These three angular 
separations are then used to determine the position vector with respect to the 
analog of the Local Group in the simulation box. Having found the direction of 
NGC 4993 in the simulation, we compute the radial projection of peculiar 
velocities of dark-matter particles found in a narrow light cone. 
Figure~\ref{pec} shows the resulting projected peculiar velocity as a function 
of the recessional velocity measured in the reference frame of the CMB 
with respect to an observer located in the Local Group. For the sake 
of readability, we downsampled from the selected dark-matter particles. 
The peculiar velocities result primarily from the proximity of the Great 
Attractor ($49\deg$ from NGC 4993).

Using a $\pm250$~km~s$^{-1}$ velocity range centered at the measured CMB rest-frame
velocity of NGC 4993, i.e., $v_{\rm CMB}=3231$ \kms, we find a mean peculiar 
velocity $v_{\rm pec}=307$~km~s$^{-1}$ and an rms $\sigma_{\rm pec} =
230$~km~s$^{-1}$. The adopted range corresponds to the smallest scale 
constrained by the observational data ($\sim5 h^{-1} {\rm Mpc}$).

Combining $\sigma_{\rm pec}$ with the error in group velocity the final cosmic 
velocity is $v_{\rm cosmic}=2924\pm 236$ \kms.

\subsection{Hubble Distance}

To a good approximation, Hubble's law gives the luminosity distance at low redshift as
\begin{equation}
D_z=H_0^{-1} c z_{\rm cosmic}\left ( 1+\frac{1-q_0}{2} z_{\rm cosmic} \right ),
\end{equation}
with $q_0=-0.53$ for standard cosmological parameters. For a local Hubble constant 
of $H_0=73.24\pm 1.74$ \kms Mpc$^{-1}$ \citep{2016ApJ...826...56R} we obtain 
$D_z=40.4\pm 3.4$ Mpc. The quoted uncertainty accounts for 
the limited depth of the NGC 4993 group of galaxies (Fig.~\ref{group}), which is
less than 1 $\deg$, corresponding to 0.6 Mpc.

We note that adopting a mean heliocentric velocity of 2995 \kms\ for the NGC 4993 group 
\citep{2017ApJ...843...16K} would increase $v_{\rm cosmic}$ by 74 \kms\ and hence the 
inferred distance by 1.0 Mpc.

\section{FP Distance}

The FP is a fairly tight relation between radius, surface brightness, and velocity 
dispersion for bulge-dominated galaxies \citep{1987ApJ...313...59D}. It is widely 
used as a distance indicator for early-type galaxies. The FP that we consider is
\begin{equation}
\log R_e + \alpha \log \sigma + \beta \log \left <I_r\right>_e + \gamma,
\end{equation}
where $R_e$ is the effective radius measured in arcseconds, $\sigma$ is the 
velocity dispersion in \kms, $\left <I_r\right>_e$ is the mean intensity inside 
the effective radius measured in L$_\odot$ pc$^{-2}$, and $\gamma$ is the 
distance-dependent zero point of the relation. 

The global values adopted by \citet{1996MNRAS.280..167J} are $\alpha=1.24$ and 
$\beta=0.82$ in Gunn $r$. We adopt $r = {\rm F606W}+0.04$ mag
\citep{1995PASP..107..945F,2005PASP..117.1049S} for E/S0 galaxies. The surface 
brightness is 
\begin{align*}
\log \left <I_r\right>_e =& -0.4 [ {\rm F606W}(<R_e)+0.04-0.29+2.5\log \pi \\
&+ 5 \log R_e -26.40-10\log(1+z)-2.5\log(1+z) ],
\end{align*}
\citep{1997ApJ...482...68H}
which transforms into Gunn $r$ and corrects for Galactic extinction
\citep{2011ApJ...737..103S}, cosmological surface brightness dimming, and 
spectral bandwidth ($k$-correction), with $z=z_{\rm cosmic}$.

In principle, FP distances may be affected by the observed magnitude range, 
morphological makeup of the sample (E-to-S0 ratio), and environment.
According to \citet{1996MNRAS.280..167J}, their version of the FP excludes 
biases above 1\% due to magnitude and morphological selection.

\subsection{FP Parameters}

\subsubsection{Photometric Properties}

The {\it HST} F606W image is fit to a 2D version of the S\'ersic function, 
\begin{equation}
I(R) = I_{e} \exp \left\{ -b_n  \left[ \left(\frac{R}{R_e} \right)^{\frac{1}{n}} 
- 1  \right]  \right\},
\end{equation}
where $R_e$ is the half-light (effective) radius, $I_e$ is the intensity at 
$R_e$, and $n$ is the fitted S\'ersic index (which is uniquely related to the 
coefficient $b_n$).

The fit was performed using the Astropy package \\
\texttt{FittingWithOutlierRemoval} 
\citep{astropy}.
The function uses sigma clipping to filter out bad data points and 
iterates $n_\textrm{iter}$ times over the data. Data deviating more than 
$n_{\sigma}$ times the standard deviation of surrounding data points 
are removed in each iteration. The filtered data are then fitted with the S\'ersic function.

The fitting algorithm was applied to a range of different combinations of 
$n_\textrm{iter}$ and $n_{\sigma}$. Fitting parameters stabilize after 15 iterations for all 
$n_{\sigma}$ values, so $n_\textrm{iter}= 15$ is used for the fit. The final $n_{\sigma}$ was
chosen for values excluding outliers and including most of the light from 
the center of the galaxy. This worked well for $n_{\sigma}$ values between 20 and 40.

\begin{figure*}[h]
    \centering
    \includegraphics[width=18cm,angle=0]{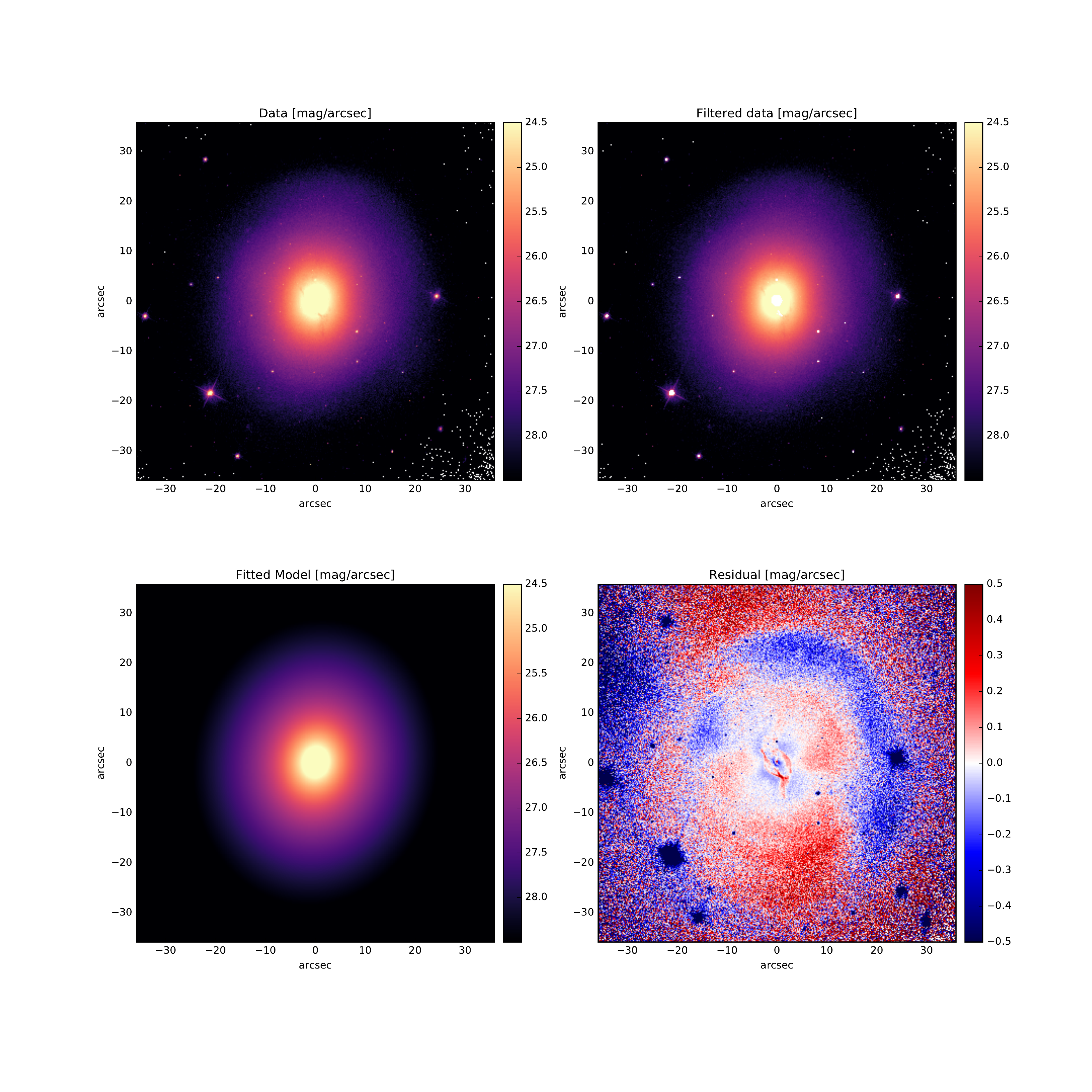}
\caption{Fits to the F606W \hst\ image of the host galaxy of GW170817. The top
left panel shows the data, the top right panel shows the masked data, 
the bottom left panel shows the fitted model, and the bottom right panel shows the 
residuals between the data and the fit. North is up, and east is to the left.
}
\label{fits}
\end{figure*}

An example of a fit to the F606W image is shown in Figure~\ref{fits}. Inside an 
effective radius of 15\farcs5 we find F606W = 12.99 mag. The S\'ersic index is
$n\approx 3.8$. A detailed discussion of the properties of NGC 4993 is presented 
in \citet{levan17}.

\subsubsection{Velocity Dispersion}

The velocity dispersion is measured in an aperture equivalent to 3\farcs4
at Coma \citep{1996MNRAS.280..167J}, i.e., in a diameter of about 8\farcs5 at 
the distance of NGC 4993. Rather than using aperture corrections from a measurement 
of the central velocity dispersion, we measure the luminosity-weighted velocity 
dispersion directly from our MUSE data to be $\sigma = 171\pm2$ \kms\ 
\citep[see also Fig.~3 of][]{levan17}. This method allows us to tie our
FP directly to that of \citet{1996MNRAS.280..167J}, thus eliminating
systematic uncertainties related to the approach adopted in determining the
velocity dispersion.

\subsection{FP Distance}
The parameters used to determine the FP distance are summarized in Table~\ref{fp}.
Calibrating the zero point of the FP to the Leo I group we can infer the distance 
to NGC 4993 as
\begin{equation}
D=10^{-\log R_e + 1.24 \log \sigma - 0.82 \left <I_r\right>_e + 2.194}
(1+z_{\rm cosmic})^2\ \rm Mpc
\end{equation}
\citep{1997ApJ...482...68H}, where the last term corrects the angular diameter 
distance to luminosity distance. The resulting distance is 44.0 Mpc.

\begin{table}[ht]
\caption{Redshift and Fundamental Plane parameters of NGC 4993}
\begin{center}
\begin{tabular}{lllll}
\hline
$z_{\rm cosmic}$ & $R_e$ (arcsec) 
& $\log \left <I_r\right>_e$ (L$_\odot$/pc$^2$) & $\sigma$ (\kms) \\
\hline
0.0098 & $15.5\pm1.5$ & $2.61\pm0.06$ & $171\pm 2$ \\
\hline
\end{tabular}
\end{center}
\label{fp}
\end{table}

\subsection{Uncertainties}

The measurement of the effective radius is subject to systematic uncertainties due 
to the nuclear spiral arms, residuals at larger radius, and difficulty in 
determining the true background level. We estimate the uncertainty in $R_e$ is 
about 1\farcs5. The combination of $R_e$ and $\left <I_r\right>_e$ in Equation (1) 
is highly degenerate \citep{1996MNRAS.280..167J} and so the FP distance is not very 
sensitive to the uncertainty in $R_e$. The measurement of $\sigma$ required no 
aperture correction since we integrate over $R_e$ in the IFU data directly. We 
estimate that the uncertainty in $\sigma$ is about 2 \kms. The uncertainties in 
$R_e$ and $\sigma$ lead to $\sim 2.1\%$ and $\sim 1.4\%$ uncertainties in the 
distance, in total 2.5\% in the distance. Therefore, the observational uncertainties 
are negligible compared to the intrinsic scatter in the FP that amounts to 17\% 
\citep{1996MNRAS.280..167J} which we adopt as the uncertainty in the FP distance to 
NGC 4993, i.e., $D_{\rm FP} = 44.0\pm7.5$ Mpc.

\section{Discussion}

\begin{figure}[h]
    \centering
    \includegraphics[width=8.8cm,angle=0]{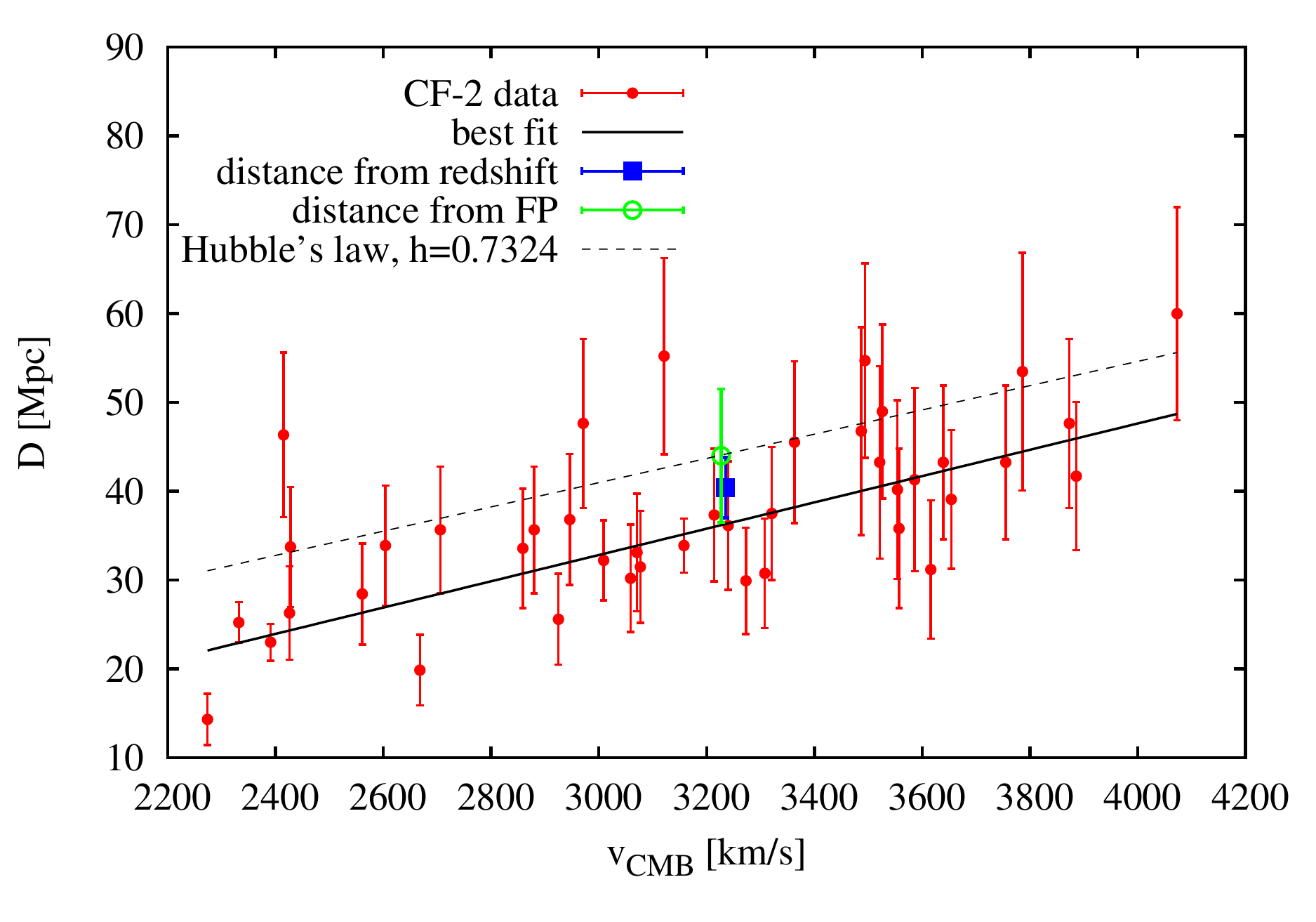}
\caption{Distance vs.\ recession velocity in the CMB frame for Cosmicflows-2 
galaxies. The best-fitting relation is consistent with the distance derived for 
NGC 4993 from the redshift and FP methods within the errors.}
\label{cf2}
\end{figure}

The two distance estimates obtained in this Letter are independent. One is based on 
a redshift and an assumed Hubble constant, and the other is based on a direct 
measurement of the distance. We can therefore combine the two distances to yield 
$D_{\rm NGC 4993}= 41.0\pm 3.1$ Mpc. This is consistent with the independent 
distance obtained from the gravitational-wave signal, $D_{\rm GW}= 
43.8^{+2.9}_{-6.9}$ Mpc \citep{lvchubble}.

As a sanity check, we can compare these estimates to a slightly different approach. 
Figure~\ref{cf2} presents an analysis of the Cosmicflows-2 catalog 
\citep{2013AJ....146...86T}. We took all distances in a range $\pm 1000$ \kms\ 
around $v_{\rm CMB}$ of NGC 4993 and 10$\deg$ around its position. From a fitted 
linear model with intrinsic scatter, the estimated distance for NGC 4993 is 
$D_{\rm CF-2}=36.2\pm 2.8$ Mpc (the error includes intrinsic scatter). We tried 
narrower cuts in $v_{\rm CMB}$ and the position, and the estimate is stable to such 
variations. This approach is not entirely independent from the other electromagnetic 
distances derived here, so we do not include it in our final value.

In Section 3, we used the \citet{2016ApJ...826...56R} value of the Hubble constant, 
$H_0=73.24\pm 1.74$ \kms\ Mpc$^{-1}$. \citet{2016A&A...594A..13P} instead find 
$H_0=67.8\pm 0.9$ \kms\ Mpc$^{-1}$ in a $\Lambda$CDM cosmology. Using the Planck 
value for $H_0$ would increase our best estimate of the luminosity distance from 
40.4 Mpc to 43.7 Mpc, a shift by $\sim 1$ standard deviation. This sensitivity to $H_0$ 
again points to the utility of these high-accuracy redshift measurements coupled with 
independent gravitational-wave distance measurements in the determination of $H_0$.

Other powerful distance indicators to distant early-type galaxies include surface 
brightness fluctuations (SBFs) and the globular cluster luminosity function (GCLF). 
Both methods are preferentially used at distances smaller than that of NGC 4993. For 
example, the GCLF peaks at $M_V\sim -7.5$ mag \citep{2012Ap&SS.341..195R}, i.e., at 
around F606W $\sim25.5$ at the distance of NGC 4993, about a magnitude shallower than 
the 2$\sigma$ detection limit for point sources in the \hst\ image \citep{levan17}. 
SBFs will be reported elsewhere. 

\section{Summary and Conclusions}

\begin{table}[ht]
\caption{Distance estimates to NGC 4993}
\begin{center}
\begin{tabular}{ll}
\hline
Method & Distance (Mpc) \\
\hline
Redshift & $40.4\pm3.4$ \\
Fundamental Plane & $44.0\pm7.5$ \\
Combined Electromagnetic ($z$+FP) & $41.0\pm3.1$ \\
\hline
Gravitational Waves & $43.8^{+2.9}_{-6.9}$ \\
\hline
\end{tabular}
\end{center}
\label{distance}
\end{table}

We have obtained a new estimate of the distance to NGC 4993 based on two methods.
(1) From a new measurement of its redshift with MUSE/VLT and an estimate of its 
peculiar velocity using CLUES constrained cosmological simulations we obtain 
$D_z=40.4\pm 3.4$ Mpc. 
(2) From new measurements of its velocity dispersion (with MUSE/VLT) and effective 
radius (\hst; Table~\ref{fp}) we obtain an FP distance of $D_{\rm FP}=44.0\pm 7.5$ Mpc. 
Combined, these result in an `electromagnetic' distance of 
$D_{\rm NGC 4993}=41.0\pm 3.1$ Mpc. This compares well with the independent 
gravitational-wave distance of $D_{\rm GW}=43.8^{+2.9}_{-6.9}$ Mpc. These results 
are summarized in Table~\ref{distance}.

The consistency between the electromagnetic and gravi\-tational-wave distances lends 
credence to both avenues of determining distances and confirms that GW170817 
occurred in NGC 4993.

There are at least two different ways in which the electromagnetically measured 
redshift to the host can lead to improved inference from gravitational-wave data.  
We can assume, as in Section 3.4, that the Hubble constant $H_0$ is known from other 
observations in order to provide better constraints on the distance. This distance 
estimate can then be fed in to gravitational-wave data analysis as a much tighter 
distance prior than the isotropic-in-volume prior $p(D) \propto D^2$ 
\citep{2015PhRvD..91d2003V}. Better distance constraints can help break some of the 
strong correlations between parameters in the gravitational-wave signal, 
particularly the distance--inclination degeneracy 
\citep{1994PhRvD..49.2658C,2013PhRvD..88f2001A,lvchubble}.
Tighter constraints on the inclination angle of the binary's orbit relative to the 
line of sight can in turn aid in the interpretation of the  electromagnetic 
transient. Alternatively, the electromagnetic redshift measurement and the 
gravitational-wave distance measurement can be combined to achieve an independent 
estimate of the Hubble constant, insensitive to the potential systematics of 
electromagnetic distance estimates \citep{1986Natur.323..310S,lvchubble}.

The prospects for improving the electromagnetic distance to NGC 4993 are good. Future 
FP studies will benefit from studies of larger samples with MUSE/VLT, which will 
allow disentangling different velocity components in galaxies and perhaps a refined 
FP indicator. The uncertainties in the peculiar velocity will also diminish as we 
obtain a better understanding of cosmic flows, e.g., along the avenue presented in 
this work. Improved photometric parameters and extension to SBF and GCLF distance 
indicators are within reach with the {\it James Webb Space Telescope}.

\acknowledgments
R.W. thanks Vinicius Ara\'ujo Barbosa de Lima for his help in handling the simulation 
data.
We thank the LIGO and Virgo Scientific Collaborations for sharing information prior 
to publication and for helpful comments on the manuscript.
We thank the anonymous referee for a swift and insightful review.
This work was supported by a VILLUM FONDEN Investigator grant to J.H. 
(project number 16599).
A.J.L. is supported by STFC and the ERC (grant \#725246).  
J.D.L. gratefully acknowledges support from STFC (ST/P000495/1).
I.M. acknowledges partial support from STFC.
C.G. acknowledges support from the Carlsberg Foundation.
This work is based in part on observations collected at the European Organisation 
for Astronomical Research in the Southern Hemisphere under ESO programme 099.D-0668.

\facilities{Hubble Space Telescope, Very Large Telescope} 
\software{{\sc astropy} \citep{astropy}}


\begin{thebibliography}{}
\expandafter\ifx\csname natexlab\endcsname\relax\def\natexlab#1{#1}\fi
\providecommand{\url}[1]{\href{#1}{#1}}

\bibitem[{{Aasi} {et~al.}(2013){Aasi}, {Abadie}, {Abbott}, {Abbott}, {Abbott},
  {Abernathy}, {Accadia}, {Acernese}, {Adams}, {Adams}, {Addesso}, {Adhikari},
  {Affeldt}, {Agathos}, {Agatsuma}, {Ajith}, {Allen}, {Allocca}, {Amador
  Ceron}, {Amariutei}, {Anderson}, {Anderson}, {Arai}, {Araya}, {Ast}, {Aston},
  {Astone}, {Atkinson}, {Aufmuth}, {Aulbert}, {Aylott}, {Babak}, {Baker},
  {Ballardin}, {Ballmer}, {Bao}, {Barayoga}, {Barker}, {Barone}, {Barr},
  {Barsotti}, {Barsuglia}, {Barton}, {Bartos}, {Bassiri}, {Bastarrika},
  {Basti}, {Batch}, {Bauchrowitz}, {Bauer}, {Bebronne}, {Beck}, {Behnke},
  {Bejger}, {Beker}, {Bell}, {Bell}, {Belopolski}, {Benacquista}, {Berliner},
  {Bertolini}, {Betzwieser}, {Beveridge}, {Beyersdorf}, {Bhadbade}, {Bilenko},
  {Billingsley}, {Birch}, {Biswas}, {Bitossi}, {Bizouard}, {Black},
  {Blackburn}, {Blackburn}, {Blair}, {Bland}, {Blom}, {Bock}, {Bodiya},
  {Bogan}, {Bond}, {Bondarescu}, {Bondu}, {Bonelli}, {Bonnand}, {Bork}, {Born},
  {Boschi}, {Bose}, {Bosi}, {Bouhou}, {Braccini}, {Bradaschia}, {Brady},
  {Braginsky}, {Branchesi}, {Brau}, {Breyer}, {Briant}, {Bridges}, {Brillet},
  {Brinkmann}, {Brisson}, {Britzger}, {Brooks}, {Brown}, {Bulik}, {Bulten},
  {Buonanno}, {Burguet{\ndash}Castell}, {Buskulic}, {Buy}, {Byer}, {Cadonati},
  {Cagnoli}, {Calloni}, {Camp}, {Campsie}, {Cannon}, {Canuel}, {Cao}, {Capano},
  {Carbognani}, {Carbone}, {Caride}, {Caudill}, {Cavagli{\`a}}, {Cavalier},
  {Cavalieri}, {Cella}, {Cepeda}, {Cesarini}, {Chalermsongsak}, {Charlton},
  {Chassande-Mottin}, {Chen}, {Chen}, {Chen}, {Chincarini}, {Chiummo}, {Cho},
  {Chow}, {Christensen}, {Chua}, {Chung}, {Chung}, {Ciani}, {Clara}, {Clark},
  {Clark}, {Clayton}, {Cleva}, {Coccia}, {Cohadon}, {Colacino}, {Colla},
  {Colombini}, {Conte}, {Conte}, {Cook}, {Corbitt}, {Cordier}, {Cornish},
  {Corsi}, {Costa}, {Coughlin}, {Coulon}, {Couvares}, {Coward}, {Cowart},
  {Coyne}, {Creighton}, {Creighton}, {Cruise}, {Cumming}, {Cunningham},
  {Cuoco}, {Cutler}, {Dahl}, {Damjanic}, {Danilishin}, {D'Antonio}, {Danzmann},
  {Dattilo}, {Daudert}, {Daveloza}, {Davier}, {Daw}, {Dayanga}, {De Rosa},
  {DeBra}, {Debreczeni}, {Degallaix}, {Del Pozzo}, {Dent}, {Dergachev},
  {DeRosa}, {Dhurandhar}, {Di Fiore}, {Di Lieto}, {Di Palma}, {Di Paolo
  Emilio}, {Di Virgilio}, {D{\'{\i}}az}, {Dietz}, {Donovan}, {Dooley},
  {Doravari}, {Dorsher}, {Drago}, {Drever}, {Driggers}, {Du}, {Dumas}, {Dwyer},
  {Eberle}, {Edgar}, {Edwards}, {Effler}, {Ehrens}, {Endr{\H o}czi}, {Engel},
  {Etzel}, {Evans}, {Evans}, {Evans}, {Factourovich}, {Fafone}, {Fairhurst},
  {Farr}, {Farr}, {Favata}, \& et~al.}]{2013PhRvD..88f2001A}
{Aasi}, J., {Abadie}, J., {Abbott}, B.~P., {et~al.} 2013, \prd, 88, 062001

\bibitem[Abbott et al.(2017a)]{lvcdisc17} 
Abbott, B. P., Abbott, R., Abbott, T. D., et al. 2017a, PhRvL, https://doi.org/
10.1103/PhysRevLett.119.161101

\bibitem[Abbott et al.(2017b)]{lvccapstone}
Abbott, B. P., Abbott, R., Abbot, T. D., et al. 2017b, ApJL, https://doi.org/
10.3847/2041-8213/aa91c9

\bibitem[Abbott et al.(2017c)]{lvchubble}
Abbott, B. P., Abbott, R., Abbott, T. D., et al. 2017c, Natur, https://doi.org/
10.1038/nature24471

\bibitem[Astropy Collaboration(2013)]{astropy} Astropy collaboration,\ 2013, \aap, 558, 33

\bibitem[Coulter et al.(2017)]{coulter17} Coulter, D.~A., Foley, R. J., Kilpatrick, C. D., et al., Sci,
http://doi.org/10.1126/science.aap9811

\bibitem[{{Crook} {et~al.}(2007){Crook}, {Huchra}, {Martimbeau}, {Masters},
  {Jarrett}, \& {Macri}}]{2007ApJ...655..790C}
{Crook}, A.~C., {Huchra}, J.~P., {Martimbeau}, N., {et~al.} 2007, \apj, 655,
  790

\bibitem[{{Cutler} \& {Flanagan}(1994)}]{1994PhRvD..49.2658C}
{Cutler}, C., \& {Flanagan}, {\'E}.~E. 1994, \prd, 49, 2658

\bibitem[{{Djorgovski} \& {Davis}(1987)}]{1987ApJ...313...59D}
{Djorgovski}, S., \& {Davis}, M. 1987, \apj, 313, 59

\bibitem[{{Fukugita} {et~al.}(1995){Fukugita}, {Shimasaku}, \&
  {Ichikawa}}]{1995PASP..107..945F}
{Fukugita}, M., {Shimasaku}, K., \& {Ichikawa}, T. 1995, \pasp, 107, 945

\bibitem[{{Gottloeber} {et~al.}(2010){Gottloeber}, {Hoffman}, \&
  {Yepes}}]{2010arXiv1005.2687G}
{Gottloeber}, S., {Hoffman}, Y., \& {Yepes}, G. 2010,  arXiv:1005.2687 

\bibitem[{{Hinshaw} {et~al.}(2009){Hinshaw}, {Weiland}, {Hill}, {Odegard},
  {Larson}, {Bennett}, {Dunkley}, {Gold}, {Greason}, {Jarosik}, {Komatsu},
  {Nolta}, {Page}, {Spergel}, {Wollack}, {Halpern}, {Kogut}, {Limon}, {Meyer},
  {Tucker}, \& {Wright}}]{2009ApJS..180..225H}
{Hinshaw}, G., {Weiland}, J.~L., {Hill}, R.~S., {et~al.} 2009, \apjs, 180, 225

\bibitem[{{Hjorth} \& {Tanvir}(1997)}]{1997ApJ...482...68H}
{Hjorth}, J., \& {Tanvir}, N.~R. 1997, \apj, 482, 68

\bibitem[{{Jorgensen} {et~al.}(1996){Jorgensen}, {Franx}, \&
  {Kjaergaard}}]{1996MNRAS.280..167J}
{Jorgensen}, I., {Franx}, M., \& {Kjaergaard}, P. 1996, \mnras, 280, 167

\bibitem[{{Kourkchi} \& {Tully}(2017)}]{2017ApJ...843...16K}
{Kourkchi}, E., \& {Tully}, R.~B. 2017, \apj, 843, 16

\bibitem[Levan et al.(2017)]{levan17} Levan, A.~L., Lyman, J. D., Tanvir, N. R., et al.\
2017, ApJL, https://doi.org/10.3847/2041-8213/aa905f

\bibitem[{{Li} {et~al.}(2014){Li}, {Hjorth}, \& {Wojtak}}]{2014ApJ...796L...4L}
{Li}, X., {Hjorth}, J., \& {Wojtak}, R. 2014, \apjl, 796, L4

\bibitem[{{Makarov} \& {Karachentsev}(2011)}]{2011MNRAS.412.2498M}
{Makarov}, D., \& {Karachentsev}, I. 2011, \mnras, 412, 2498

\bibitem[Pian et al.(2017)]{pian17} Pian, E., D'Avanzo, P., Benetti, S., et al.\ 2017,
Natur, http://doi.org/10.1038/nature24298

\bibitem[{{Planck Collaboration}(2016){Planck Collaboration}, {Ade},
  {Aghanim}, {Arnaud}, {Ashdown}, {Aumont}, {Baccigalupi}, {Banday},
  {Barreiro}, {Bartlett}, {Bartolo}, {Battaner}, {Battye}, {Benabed},
  {Beno{\^i}t}, {Benoit-L{\'e}vy}, {Bernard}, {Bersanelli}, {Bielewicz},
  {Bock}, {Bonaldi}, {Bonavera}, {Bond}, {Borrill}, {Bouchet}, {Boulanger},
  {Bucher}, {Burigana}, {Butler}, {Calabrese}, {Cardoso}, {Catalano},
  {Challinor}, {Chamballu}, {Chary}, {Chiang}, {Chluba}, {Christensen},
  {Church}, {Clements}, {Colombi}, {Colombo}, {Combet}, {Coulais}, {Crill},
  {Curto}, {Cuttaia}, {Danese}, {Davies}, {Davis}, {de Bernardis}, {de Rosa},
  {de Zotti}, {Delabrouille}, {D{\'e}sert}, {Di Valentino}, {Dickinson},
  {Diego}, {Dolag}, {Dole}, {Donzelli}, {Dor{\'e}}, {Douspis}, {Ducout},
  {Dunkley}, {Dupac}, {Efstathiou}, {Elsner}, {En{\ss}lin}, {Eriksen},
  {Farhang}, {Fergusson}, {Finelli}, {Forni}, {Frailis}, {Fraisse},
  {Franceschi}, {Frejsel}, {Galeotta}, {Galli}, {Ganga}, {Gauthier}, {Gerbino},
  {Ghosh}, {Giard}, {Giraud-H{\'e}raud}, {Giusarma}, {Gjerl{\o}w},
  {Gonz{\'a}lez-Nuevo}, {G{\'o}rski}, {Gratton}, {Gregorio}, {Gruppuso},
  {Gudmundsson}, {Hamann}, {Hansen}, {Hanson}, {Harrison}, {Helou},
  {Henrot-Versill{\'e}}, {Hern{\'a}ndez-Monteagudo}, {Herranz}, {Hildebrandt},
  {Hivon}, {Hobson}, {Holmes}, {Hornstrup}, {Hovest}, {Huang}, {Huffenberger},
  {Hurier}, {Jaffe}, {Jaffe}, {Jones}, {Juvela}, {Keih{\"a}nen}, {Keskitalo},
  {Kisner}, {Kneissl}, {Knoche}, {Knox}, {Kunz}, {Kurki-Suonio}, {Lagache},
  {L{\"a}hteenm{\"a}ki}, {Lamarre}, {Lasenby}, {Lattanzi}, {Lawrence}, {Leahy},
  {Leonardi}, {Lesgourgues}, {Levrier}, {Lewis}, {Liguori}, {Lilje},
  {Linden-V{\o}rnle}, {L{\'o}pez-Caniego}, {Lubin}, {Mac{\'{\i}}as-P{\'e}rez},
  {Maggio}, {Maino}, {Mandolesi}, {Mangilli}, {Marchini}, {Maris}, {Martin},
  {Martinelli}, {Mart{\'{\i}}nez-Gonz{\'a}lez}, {Masi}, {Matarrese}, {McGehee},
  {Meinhold}, {Melchiorri}, {Melin}, {Mendes}, {Mennella}, {Migliaccio},
  {Millea}, {Mitra}, {Miville-Desch{\^e}nes}, {Moneti}, {Montier}, {Morgante},
  {Mortlock}, {Moss}, {Munshi}, {Murphy}, {Naselsky}, {Nati}, {Natoli},
  {Netterfield}, {N{\o}rgaard-Nielsen}, {Noviello}, {Novikov}, {Novikov},
  {Oxborrow}, {Paci}, {Pagano}, {Pajot}, {Paladini}, {Paoletti}, {Partridge},
  {Pasian}, {Patanchon}, {Pearson}, {Perdereau}, {Perotto}, {Perrotta},
  {Pettorino}, {Piacentini}, {Piat}, {Pierpaoli}, {Pietrobon}, {Plaszczynski},
  {Pointecouteau}, {Polenta}, {Popa}, {Pratt}, {Pr{\'e}zeau}, {Prunet},
  {Puget}, {Rachen}, {Reach}, {Rebolo}, {Reinecke}, {Remazeilles}, {Renault},
  {Renzi}, {Ristorcelli}, {Rocha}, {Rosset}, {Rossetti}, {Roudier},
  {Rouill{\'e} d'Orfeuil}, {Rowan-Robinson}, {Rubi{\~n}o-Mart{\'{\i}}n},
  {Rusholme}, {Said}, {Salvatelli}, {Salvati}, {Sandri}, {Santos},
  {Savelainen}, {Savini}, {Scott}, {Seiffert}, {Serra}, {Shellard}, {Spencer},
  {Spinelli}, {Stolyarov}, \& et~al.}]{2016A&A...594A..13P}
{Planck Collaboration}, {Ade}, P.~A.~R., {Aghanim}, N., {et~al.} 2016, \aap,
  594, A13
  
\bibitem[{{Rejkuba}(2012)}]{2012Ap&SS.341..195R}
{Rejkuba}, M. 2012, \apss, 341, 195

\bibitem[{{Riess} {et~al.}(2016){Riess}, {Macri}, {Hoffmann}, {Scolnic},
  {Casertano}, {Filippenko}, {Tucker}, {Reid}, {Jones}, {Silverman},
  {Chornock}, {Challis}, {Yuan}, {Brown}, \& {Foley}}]{2016ApJ...826...56R}
{Riess}, A.~G., {Macri}, L.~M., {Hoffmann}, S.~L., {et~al.} 2016, \apj, 826, 56

\bibitem[{{Schlafly} \& {Finkbeiner}(2011)}]{2011ApJ...737..103S}
{Schlafly}, E.~F., \& {Finkbeiner}, D.~P. 2011, \apj, 737, 103

\bibitem[{{Schutz}(1986)}]{1986Natur.323..310S}
{Schutz}, B.~F. 1986, \nat, 323, 310

\bibitem[{{Sirianni} {et~al.}(2005){Sirianni}, {Jee}, {Ben{\'{\i}}tez},
  {Blakeslee}, {Martel}, {Meurer}, {Clampin}, {De Marchi}, {Ford}, {Gilliland},
  {Hartig}, {Illingworth}, {Mack}, \& {McCann}}]{2005PASP..117.1049S}
{Sirianni}, M., {Jee}, M.~J., {Ben{\'{\i}}tez}, N., {et~al.} 2005, \pasp, 117,
  1049

\bibitem[{{Sorce} {et~al.}(2014){Sorce}, {Tully}, {Courtois}, {Jarrett},
  {Neill}, \& {Shaya}}]{2014MNRAS.444..527S}
{Sorce}, J.~G., {Tully}, R.~B., {Courtois}, H.~M., {et~al.} 2014, \mnras, 444,
  527

\bibitem[{{Springob} {et~al.}(2014){Springob}, {Magoulas}, {Colless}, {Mould},
  {Erdo{\u g}du}, {Jones}, {Lucey}, {Campbell}, \&
  {Fluke}}]{2014MNRAS.445.2677S}
{Springob}, C.~M., {Magoulas}, C., {Colless}, M., {et~al.} 2014, \mnras, 445,
  2677

\bibitem[{{Springob} {et~al.}(2009){Springob}, {Masters}, {Haynes},
  {Giovanelli}, \& {Marinoni}}]{2009ApJS..182..474S}
{Springob}, C.~M., {Masters}, K.~L., {Haynes}, M.~P., {Giovanelli}, R., \&
  {Marinoni}, C. 2009, \apjs, 182, 474

\bibitem[Tanvir et al.(2017)]{tanvir17} Tanvir, N.~R., Levan, A. J., Gonz\'alez-Fern\'andez, C.,
et al.\ 2017, ApJL, https://doi.org/10.3847/2041-8213/aa90b6

\bibitem[{{Tully} {et~al.}(2013){Tully}, {Courtois}, {Dolphin}, {Fisher},
  {H{\'e}raudeau}, {Jacobs}, {Karachentsev}, {Makarov}, {Makarova},
  {Mitronova}, {Rizzi}, {Shaya}, {Sorce}, \& {Wu}}]{2013AJ....146...86T}
{Tully}, R.~B., {Courtois}, H.~M., {Dolphin}, A.~E., {et~al.} 2013, \aj, 146,
  86

\bibitem[{{Tully} {et~al.}(2016){Tully}, {Courtois}, \&
  {Sorce}}]{2016AJ....152...50T}
{Tully}, R.~B., {Courtois}, H.~M., \& {Sorce}, J.~G. 2016, \aj, 152, 50

\bibitem[{{Tully} {et~al.}(2009){Tully}, {Rizzi}, {Shaya}, {Courtois},
  {Makarov}, \& {Jacobs}}]{2009AJ....138..323T}
{Tully}, R.~B., {Rizzi}, L., {Shaya}, E.~J., {et~al.} 2009, \aj, 138, 323

\bibitem[{{Veitch} {et~al.}(2015){Veitch}, {Raymond}, {Farr}, {Farr}, {Graff},
  {Vitale}, {Aylott}, {Blackburn}, {Christensen}, {Coughlin}, {Del Pozzo},
  {Feroz}, {Gair}, {Haster}, {Kalogera}, {Littenberg}, {Mandel},
  {O'Shaughnessy}, {Pitkin}, {Rodriguez}, {R{\"o}ver}, {Sidery}, {Smith}, {Van
  Der Sluys}, {Vecchio}, {Vousden}, \& {Wade}}]{2015PhRvD..91d2003V}
{Veitch}, J., {Raymond}, V., {Farr}, B., {et~al.} 2015, \prd, 91, 042003

\bibitem[{{Willick} {et~al.}(1997){Willick}, {Courteau}, {Faber}, {Burstein},
  {Dekel}, \& {Strauss}}]{1997ApJS..109..333W}
{Willick}, J.~A., {Courteau}, S., {Faber}, S.~M., {et~al.} 1997, \apjs, 109,
  333

\end{thebibliography}
\end{document}